\documentstyle{article}
\begin{document}
\rightline{IC/93/288}
\rightline{UTS-DFT-93-22}
\vspace*{2cm}
\Large
\begin{center}
{\bf PROBING THREE-BOSON ANOMALOUS COUPLINGS IN $e^+e^-\to W^+W^-$
AT FUTURE LINEAR $e^+e^-$ COLLIDERS WITH POLARIZED BEAMS}
\end{center}
\large
\vspace*{1cm}
\begin{center}
A.A.Likhoded$^{a)}$, A.A.Pankov$^{b)}$
\footnote
{\large Permanent address: Gomel Polytechnical Institute,
Gomel, 246746 Belarus, CIS\\
E-mail: PANKOV\%GPI.GOMEL.BY~@~RELAY.USSR.EU.NET},
N.Paver$^{c)}$, M.V.Shevlyagin$^{a,\: d)}$ and O.P.Yushchenko$^{d)}$
\end{center}
\vspace*{1.5cm}
\begin{center}
$^{a)}$ {\it
The Branch of Institute for  Nuclear Physics, Protvino,
 142284 Russia}\\
\end{center}

\begin{center}
$^{b)}$ {\it
International Centre for Theoretical Physics, Trieste, Italy\\
and\\
Istituto Nazionale di Fisica Nucleare, Sezione di Trieste, Trieste, Italy}\\
\end{center}

\begin{center}
$^{c)}$ {\it
Dipartimento di Fisica Teorica, Universit\`a di Trieste, Trieste, Italy\\
and\\
Istituto Nazionale di Fisica Nucleare, Sezione di Trieste, Trieste, Italy}\\
\end{center}

\begin{center}
$^{d)}$ {\it Institute for High Energy Physics, Protvino,
 142284 Russia}\\
\end{center}
\vspace*{0,5cm}
\begin{center}
{\bf ABSTRACT}
\end{center}
\vspace*{0.4cm}
\noindent

We study the potential of the next-generation $e^+ e^-$ linear colliders
with longitudinally polarized beams, to restrict the values of the
anomalous trilinear couplings $WW\gamma$ and $WWZ$ from the measurement of
the process
$e^+ e^-\to W^+ W^-$. Along with initial $e^+e^-$ polarization, we account
also for the possibilities offered by
cross sections for polarized final $W$, in order to disentangle the constraints
on the various constants. The results show the essential role of the
initial beams polarization in improving the bounds obtained from the
unpolarized case.

\newpage
\large
\vspace*{4cm}
\baselineskip=18pt
\section{Introduction}

The experimental confirmation of the Standard Model (SM) is presently
limited to the sector of the interaction of fermions with vector bosons.
Another key ingredient of the SM is represented by the vector boson
self-interactions, which are a consequence of the non-abelian structure of
the electroweak symmetry, and are essential for the renormalizability of the
theory.

The precise measurement of the three-boson $WW\gamma$ and $WWZ$ couplings
is an important item in the physics programme at planned
high energy (and high luminosity) colliders [1,2].
In the SM these vertices are exactly
determined  by the $SU(2)_L\times U(1)_Y$ gauge symmetry, and therefore
their measurement gives a unique chance to test the gauge
structure of the electroweak theory. While experiments at low energy and
precision measurements at the $Z^0$ pole can provide {\it indirect}
access to these constants, only very high energy colliders, well above
the threshold for $W$-pair production, will allow {\it direct}
and unambiguous tests. Indeed, in the near future one can foresee
analyses of boson self-couplings at LEP2 [3,4] and to some extent at the
Tevatron [4-7] and HERA [8].

A new stage in precision in this field will be reached at the planned
hadron-hadron (SSC, LHC) and $e^+e^-$ linear colliders (NLC, JLC, VLEPP),
owing to the
enhanced sensitivity to deviations from the SM, in particular to anomalous
values of the gauge boson self-couplings, allowed by the significantly higher
energies of these machines.

Among the various possible reactions, where to test the trilinear gauge boson
couplings, a special role will be played by the process
\begin{equation}e^++e^-\to W^++W^-,\label{proc}\end{equation}
at $e^+e^-$ linear colliders  [1,2,9--16].  This process should be
particularly sensitive to deviations of the gauge boson couplings
from the SM, originating from
some ``new physics'' source. Such an enhancement of the sensitivity reflects
the lack of compensation among the individual, $s$-diverging contributions to
the SM
cross section, corresponding at the Born level to
$\gamma$, $\nu$ and $Z$ exchange diagrams and their interferences. Instead,
in the absence of new physics, such a gauge cancellation exactly occurs,
and consequently the SM cross section decreases with $\sqrt s$ [17,18].

Indeed, in the specific case of modifications of the $\gamma$- and $Z$-mediated
amplitudes, induced by anomalous values of the trilinear gauge boson couplings
$WW\gamma$ and $WWZ$, such that:
\begin{equation}A_V\to(1+\epsilon_V)A_V,\end{equation}
where $V=\gamma,Z$, we can define the relative deviation of the cross section
of process (1) from the SM prediction as follows:
\begin{equation}\Delta(s)=\frac{\sigma (s,\epsilon_V)-\sigma(s)^{SM}}
{\sigma (s)^{SM}}.\label{delta}\end{equation}
Here, $\sigma$ represents the cross section (either total, or differential,
or integrated in some angular range), and
\begin{eqnarray}
\sigma(s,\epsilon_V)&\propto&\vert(1+\epsilon_{\gamma})A_{\gamma}
+A_{1\nu}+(1+\epsilon_Z)A_Z\vert^2+\vert A_{2\nu}\vert^2,\nonumber \\
\sigma^{SM}(s)&\propto&\vert A_{\gamma}+A_{1\nu}+A_Z\vert^2
+\vert A_{2\nu}\vert^2,
\label{vertex}\end{eqnarray}
where for later convenience the neutrino-exchange amplitude is split
into a part $A_{1\nu}$ interfering with the $s$-channel diagrams and a
non-interfering part $A_{2\nu}$, such that $A_{\nu}=A_{1\nu}+A_{2\nu}$.
One obtains:
\begin{equation}\Delta=\Delta_{\gamma}+\Delta_Z,\end{equation}
where
\begin{eqnarray}\Delta_\gamma=\epsilon_{\gamma}(R_{\nu\gamma}+R_{Z\gamma}
+2R_{\gamma\gamma}),
\nonumber \\
\Delta_Z=\epsilon_Z(R_{\gamma Z}+R_{\nu Z}+2R_{ZZ})
\label{delgz}\end{eqnarray}
and ($i,j=\gamma,\nu,Z$)
\begin{equation}R_{ij}=\sigma_{ij}/\sigma^{SM};\qquad
\sigma^{SM}(s)\equiv\sigma(s,\epsilon_V=0)=\sum_{i,j}
\sigma_{ij}.\label{rij}
\end{equation}
Concerning the separation of the $\nu$-exchange diagram into $A_{1\nu}$ and
$A_{2\nu}$, introducing the helicities of $W^-\ (W^+)$ as $\lambda\
(\bar\lambda)$, the amplitude $A_{2\nu}$ has
$\vert\Delta\lambda\vert=\vert\lambda-\bar\lambda\vert=2$, while all the
others have $\vert\Delta\lambda\vert=0,\ 1$. It turns out that $A_{2\nu}$ is
numerically dominant with respect to the other amplitudes, in the energy range
considered below.

In Eqs.(5--7), $\Delta$'s are determined by linear combinations of non
cancelling individually divergent contributions, and will increase, basically
like a power of $s$. In contrast, the SM cross section decreases at least
as $1/s$. Thus, if we parametrize the sensitivity of process (1) to
$\epsilon_V$ by {\it e.g.} the ratio
${\cal S}={\Delta/(\delta\sigma/\sigma)}$, with
${\delta\sigma/\sigma}$ the statistical uncertainty experimentally attainable
on the SM cross section, such a sensitivity is power-like enhanced with
increasing $\sqrt s$, even at fixed integrated luminosity.

An additional, and quite significant, improvement in the sensitivity should be
obtained
if initial $e^+e^-$ longitudinal polarizations were available, so that one
could separately measure the cross sections for both $e^-_Le^+_R$
($\sigma^{LR}$) and $e^-_Re^+_L$ ($\sigma^{RL}$), as discussed in
[19]. One
should notice that $\sigma^{RL}$ does not contain the neutrino-exchange
diagram, which dominates the SM cross section and is not modified by the
anomalous trilinear couplings. Therefore, this diagram represents a sort of
``background'' in the kind of searches discussed here. Consequently, although
${\sigma^{RL}\ll\sigma^{LR}}$ leads a much lower statistics,
in principle one can  qualitatively expect to derive stringent limits in the
$RL$ case also. In fact, in practice longitudinal polarization
will not exactly be 100\%, and for realistic values of the polarization
the determination of the $RL$ cross section from the data could be totally
obscured by the uncertainty in the polarization itself. Due to
${\sigma^{LR}\gg\sigma^{RL}}$, such an uncertainty could induce a systematic
error on $\sigma^{RL}$ much larger than the statistical error for this cross
section, and consequently the sensitivity would be diminished. However, as
it will be discussed in the sequel, one can find ``optimal'' kinematical
regions to integrate cross
sections, where this effect does not so dramatically contribute to the
uncertainty on the $RL$ cross section, and therefore the expected sensitivity
on the anomalous coupling constants
provided by $\sigma^{RL}$ qualitatively remains the same. Clearly, as in
[19], the complete analysis should combine measurements of
$\sigma^{RL}$ and $\sigma^{LR}$, in particular for the purpose of
disentangling the constraints for the different anomalous vertices.

In this regard, combined measurements of cross sections for
final polarized (longitudinal
and/or transverse) $W$'s with polarized initial beams would also be
extremely useful both to improve the sensitivity to individual anomalous
couplings and to separate the various dependences.

In this paper we will present the bounds on the anomalous three-boson
constants, which can be obtained along the lines sketched above from the
consideration of the process  $e^+e^-\to W^+W^-$ at future high-luminosity
linear ${e^+e^-}$ colliders, with CM energies of ${0.5-1\ TeV}$ and polarized
beams, assuming that also $W^+W^-$ polarizations will be measured.

Specifically, in Section 2 we will introduce the standard
parameterization of the $WW\gamma$ and $WWZ$ vertices in terms of the familiar
notation using $k_V$ and $\lambda_V$, and will briefly review the current
bounds on these parameters as well as the expectations from forthcoming
experiments. In Section 3 we analyze in details the potential of process (1)
and the role of polarization, and assess the resulting constraints on the
anomalous couplings.  Finally, Section 4 will be devoted to a discussion of
the results and to some concluding remarks.

\section{WWV vertices}

Limiting to the $C$ and $P$ invariant part of the interaction,
\footnote{\large In this paper we do not consider $C$, $P$ and $T$ violating
operators. The latter should be strongly suppressed, {\it e.g.} by
the upper bound on the neutron electric dipole moment, which implies for the
$WW\gamma$ vertex $|\tilde k_{\gamma}|,\: |\tilde \lambda_{\gamma}|\leq
O(10^{-4})$ [20]. We also neglect a possible deviation of the
Yang-Mills coupling constants $g_{WWV}$ from those predicted by the SM.} the
$WWV$ coupling, represented in Fig.1, is [9]:
\begin{equation}
{\cal L}_{eff}=-ig_{WWV}\left[W^+_{\mu\nu}W^{\mu}V^{\nu}-
W^+_{\mu}V_{\nu}W^{\mu\nu}+k_V W^+_{\mu}W_{\nu}F^{\mu\nu}+
\frac{\lambda_V}{M^2_W}W^+_{\lambda\mu}W^{\mu}_{\nu}F^{\lambda\nu}
\right].
\end{equation}
Here, $W^{\mu}$ is the $W^-$ boson field,
$W_{\mu\nu}=\partial_{\mu}W_{\nu} - \partial_{\nu}W_{\mu}$,
$F_{\mu\nu}=\partial_{\mu}V_{\nu} - \partial_{\nu}V_{\mu}$,
and the gauge coupling constants $g_{WWV}$ are, with ${V=\gamma,\ Z}$
\begin{equation}
g_{WW\gamma}=e\hskip 2pt;\qquad\quad g_{WWZ}=e\cdot cot\theta_W,
\end{equation}
where $e$ is the electron charge and ${\theta}_W$ the electroweak mixing
angle.

For the $WW\gamma$ coupling in the static limit there is a simple
interpretation of the parameters appearing in Eq.(8), namely
$g_{WW\gamma}$ defines the $W$ electric charge, while $k_{\gamma}$ and
$\lambda_{\gamma}$ are connected with the magnetic ($\mu _W$) and  electric
quadrupole ($Q_W$) moments of the $W$ boson:
\begin{equation}
\mu_W={e\over {2M_W}}(1+k_{\gamma}+\lambda_{\gamma})\quad ,\qquad
Q_W=-{e\over {M_W^2}}(k_{\gamma}-\lambda_{\gamma}).
\end{equation}
A similar interpretation holds for the parameters
$k_{Z}$ and $\lambda_Z$ in the $WWZ$ vertex.

Referring to Fig.1, in momentum space the vertex of Eq.(8) can be written
as [9]:
\begin{equation}
\Gamma_{\mu\alpha\beta }^V(q,\overline q,p)=(\bar q -q)_{\mu}
\left[(1+\frac{\lambda_V p^2}{2M_W^2})g_{\alpha\beta}-
\lambda_V\frac{p_{\alpha}
p_{\beta}}{M^2_W}\right]+
\left(p_{\beta}g_{\mu\alpha}-p_{\alpha}g_{\mu\beta})
(1+k_V +\lambda_V\right)\hskip 2pt .\end{equation}

In the SM at the tree level, $k_{V}=1$ and $\lambda_{V}=0.$
The existing limits on $k_{V}$ and $\lambda_{V}$ are rather loose, $\leq O(1)$.
Bounds on the anomalous moments can be derived from high-precision measurements
of electroweak observables which are affected by $W$--loop corrections [21,22].
These effects have been discussed for the $Z$ boson parameters and for atomic
parity violation. From low-energy data and from precision LEP~I measurements:
\begin{eqnarray}
\left|\lambda_{\gamma}\right| \leq 0.6 && \qquad \quad
\left|k_{\gamma}-1\right| \leq 1.0 \nonumber\\ *
\left|\lambda_{Z}\right| \leq 0.6 &&  -0.8\leq k_{Z}-1 \leq 0\; \nonumber
\end{eqnarray}
However, some combinations of $k_{V}$ and $\lambda_{V}$
can be restricted more tightly: for example,
$|\lambda_{\gamma}-\lambda_{Z}| \leq 0.1$ for any value of
$\lambda_{\gamma,Z}$ and  $|\lambda_{\gamma}-\lambda_{Z}| \leq 0.01$ for
$\lambda_{\gamma,Z}>0.25$ .
In perspective, from the $q\overline q \to W\gamma$ and
$q\overline q \to WZ$ processes at the Tevatron with an integrated
luminosity of 1 $fb^{-1}$, one can obtain the following bounds:
\begin{eqnarray}
\left|\lambda_{\gamma}\right| \leq 0.2 && -0.50\leq k_{\gamma}-1 \leq 0.80
\nonumber\\ *
\left|\lambda_{Z}\right| \leq 0.4 &&  -0.80\leq k_{Z}-1 \leq 0\nonumber
\end{eqnarray}
The corresponding limits at LEP II with $\sqrt{s}\simeq 200$ GeV will
be [3,4]:
\begin{eqnarray}
\left|\lambda_{\gamma}\right| \leq 0.4 && -0.14\leq k_{\gamma}-1 \leq 0.87
\nonumber\\ *
\left|\lambda_{Z}\right| \leq 0.4 &&  -0.24\leq k_{Z}-1 \leq 0\nonumber
\end{eqnarray}

For the higher energy $e^+e^-$ linear colliders,
the study of the  process $e^+e^- \to W^+W^-$  with no initial longitudinal
polarization can lead to limits on ${\vert \lambda_V\vert}$ and
${\vert k_V\vert}$ typically of the order of some {\it percent}
[1,2]. Regarding the separation of
$k_{V}$ from $\lambda_{V}$, the advantages
of using the processes $e^-\gamma \to W^-\nu$,
$\gamma\gamma\to W^+W^-$ and $e^+e^-\to\nu\overline\nu Z(\to \mu^+\mu^-)$
have been investigated in the literature [11,12,23-26]. However, for these
processes
the sensitivity to anomalous trilinear couplings turns out not to be improved
with respect to the numbers reported above. Also, restrictions similar to those
found at linear $e^+e^-$ colliders could be obtained from the complementary
studies at the hadron-hadron supercolliders [4-7].

In the next Section we are going to discuss in details the potential of
process (1) to constrain $\lambda_{V}$ and $k_V$, particularly
emphasizing the role of {\it initial}, in addition to final states
polarizations.

\section{The process $e^+e^-\to W^+W^-$ with polarization}

The $e^+ e^-$ annihilation into a $W$--pair is determined in Born
approximation by the diagrams in Fig.2. We start our discussion with the
case of initial $RL$ polarization, and with final longitudinal $W_LW_L$ or
transverse $W_TW_T$ polarizations. In the sequel these two cases for the
$W$ polarizations
will be denoted by $LL$ and $TT$, respectively. The corresponding transition
amplitudes can be written as (see Appendix):
\begin{equation}{\cal A}^{RL}_{LL}=\frac{s}{M_W^2}\ \left[\frac{3-\beta_W^2}{2}
\left(1-\chi\cdot g_Zg_R\right)+
\left(\Delta k_{\gamma}-\chi\cdot g_Zg_R\Delta k_Z\right)\right],\label{all}
\end{equation}
and
\begin{equation}{\cal A}^{RL}_{TT}=\left(1-\chi\cdot g_Zg_R\right)+
\frac{s}{2M_W^2}\
\left(\lambda_{\gamma}-\chi\cdot g_Zg_R\hskip 2pt\lambda_Z\right).\label{att}
\end{equation}
Here and in the following the notations are such that upper indices refer to
initial $e^-e^+$ longitudinal polarizations, and the lower indices indicate
the final $W^{\pm}$ longitudinal and/or transverse polarizations. In (12) and
(13), which are easily derived using Table 3.1 of Ref.[1]:
${\Delta k_V=k_V-1}$ and $\lambda_V$  are the anomalous trilinear couplings
as defined in (8);
${\beta_W=\sqrt{1-4M_W^2/s}}$; ${g_Z=\cot{\theta_W}}$;
${g_R=\tan{\theta_W}}$ is the right-handed electron coupling constant;
and finally ${\chi=s/(s-M_Z^2)}$ is the $Z$ propagator. The explicit
expressions of $\epsilon_V$ introduced in Eq.(2) can be easily obtained
for the various polarizations from Eqs.(\ref{all})--(\ref{att}). Notice that in
these equations we have not explicitly
included an angular-dependent factor $\propto\sin\theta$, common to all
s-channel helicity amplitudes.

In terms of the amplitudes (\ref{all}) and (\ref{att}), the total cross
sections, integrated over all angles, are given by:
\begin{equation}\sigma^{RL}_{LL}=\frac{\pi\alpha_{\it e.m.}^2
\beta_W^3}{6s}\vert{\cal A}^{RL}_{LL}\vert^2,\label{sll}\end{equation}
and
\begin{equation}\sigma^{RL}_{TT}=\frac{4\pi\alpha_{\it e.m.}^2\beta_W^3}
{3s}\vert{\cal A}^{RL}_{TT}\vert^2.\label{stt}\end{equation}
One can notice that $\sigma^{RL}_{LL}$ and $\sigma^{RL}_{TT}$ separately
depend on different sets of trilinear gauge boson couplings, and therefore give
independent information on $k$'s and $\lambda$'s. In contrast, the
$LT+TL$ cross section turns out to
depend on all four couplings and therefore is not so useful for
disentangling their effects. Detailed formulae and definitions are collected
in the Appendix, which can also be used to derive explicit expressions for the
deviations from the SM due to anomalous couplings, defined in
Eqs.(\ref{delgz})--(\ref{rij}),
for the different polarizations.

In Fig.3 we represent the energy behavior of $\sigma^{RL}_{LL}$ and
$\sigma^{RL}_{TT}$ in the SM, along with the cross section for unpolarized
$W$ bosons. The latter case includes the sum over the three possibilities
$LL$, $TT$ and $TL+LT$. In accordance to Eqs.(\ref{sll})--(\ref{stt}),
Fig.3 shows that
the $LL$ cross section is largely dominating for increasing energy over the
$TT$ cross section, by a factor of order $\left(s/M_W^2\right)^2$.
Consequently, sensitivities on anomalous couplings expected from statistical
arguments, denoted as ${\cal S}$ in Section 1, will have the behavior
${\cal S}_{LL}\propto s^{3/2}$ and, provided $\sigma^{RL}_{TT}$ could be
measured, ${\cal S}_{TT}\propto s^{5/2}$. Anyway, as discussed later in this
Section, although much smaller than in the $LL$ case, the $TT$ cross section
could be useful in order to constrain
the values of the couplings $\lambda_{\gamma}$ and $\lambda_Z$. Also,
observing from Fig.3 that
the $LL$ cross section is numerically close to the unpolarized one, we
obviously conclude that the latter cross section is mostly sensitive to
$\Delta k_V$, rather than to $\lambda_V$.

To derive the typical values of the bounds on anomalous couplings, that can be
derived from $e^+_Le^-_R\to W^+W^-$, we choose two possible values of the
CM energy,
referring to the planned NLC colliders [27], namely $\sqrt s=0.5\ TeV$ and
$\sqrt s=1\ TeV$, with integrated luminosities $L_{int}=50\ fb^{-1}$ and
$L_{int}=100\ fb^{-1}$, respectively. In both cases we use the channel of two
leptons plus two hadronic jets $(l\hskip 3pt\nu +j\hskip 2pt j)$ to identify
the final $W^+W^-$ state, with an efficiency of reconstruction
$\varepsilon_W=0.15$. Referring to Eq.(\ref{delta}), in this Section we
discuss the bounds which would be derived from just statistical arguments,
{\it i.e.} by demanding that
\begin{equation}\vert\Delta\vert=\frac{\vert\sigma (\Delta k_V,\lambda_V)-
\sigma^{SM}\vert}
{\sigma^{SM}}<\frac{\delta\sigma}{\sigma},\label{delv}\end{equation}
with $\delta\sigma/\sigma$ the statistical uncertainty in the specific cases.
We defer to the next Section a presentation of the bounds taking into account
also some systematic uncertainties.

Then, by solving the inequality (\ref{delv}) in the case of
$\sigma^{RL}_{LL}$, we obtain the two conditions:
\begin{equation}\vert\Delta k_{\gamma}-\chi\cdot\Delta k_Z\vert<
{\frac{1}{2}}\left(\frac{\delta\sigma}{\sigma}\hskip 2pt\vert
{\tilde{\cal A}}\vert
\right)_{LL}^{RL},\label{k1}
\end{equation}
and
\begin{equation}\vert\Delta k_{\gamma}-\chi\hskip 2pt\Delta k_Z+
2{\tilde{\cal A}}^{RL}_{LL}
\vert<
{\frac{1}{2}}\left(\frac{\delta\sigma}{\sigma}\hskip 2pt\vert{\tilde{\cal A}}
\vert\right)_{LL}^{RL},\label{k2}
\end{equation}
where ${\displaystyle
{\tilde{\cal A}}^{RL}_{LL}=\frac{3-\beta_W^2}{2}(1-\chi)}$.
Eqs.(\ref{k1}) and (\ref{k2}) are derived under the assumption that the
statistical
uncertainty $\delta\sigma/\sigma\ll1$, so that one can expand (\ref{delv})
to first order in this parameter. As seen from Fig.3, in the chosen range of
$\sqrt s$ the values of the SM cross sections are
$\sigma^{RL}_{LL}=84\ fb$ and $22\ fb$ for $\sqrt s=0.5\ TeV$
and $1\ TeV$, respectively. Correspondingly,  from (\ref{sll}),
for the right-hand sides of Eqs.(\ref{k1})--(\ref{k2}) we find the values:
\begin{equation}{\frac{1}{2}}\left(\frac{\delta\sigma}{\sigma}
\hskip 2pt\vert{\tilde{\cal A}}\vert\right)_{LL}^{RL}=
\frac{1}{\sqrt{(\sigma_0)_{LL}\varepsilon_W L_{int}}}\cdot\frac{M_W^2}{s}=
1.3\cdot 10^{-3}\ (4.6\cdot 10^{-4}),\label{atilda}\end{equation}
at the 95\% C.L. for the two values of the CM energy, with
$(\sigma_0)_{LL}=\pi\alpha_{\it e.m.}^2\beta_W^3/6s$.

In Fig.4 the bands labeled as `1' and `2' represent the allowed domains for
the anomalous constants, resulting from the inequalities above.
\footnote{\large If we
account for the imaginary part of the propagator, {\it i.e.}
$\chi\to s/(s-M_Z^2+iM_Z\Gamma_Z)$, the straight bands transform into a region
enclosed by two ellipses.} At this stage, since the bands are not limited,
we have no restrictions on $\Delta k_{\gamma}$ and $\Delta k_Z$ separately, but
only correlations between their values. To obtain separate bounds
one has to change the {\it slopes} of the bands, in such a way as to find
new bands crossing the preceding ones `1' and `2'. To this purpose,
referring to the left sides of Eqs.(\ref{k1}) and (\ref{k2}), there
are two possibilities, namely either to exploit the $s$-behavior of the $Z$
propagator $\chi$, affecting the slopes, or to change the initial
beams polarization, and consider {\it e.g.} the $LR$ cross section. Since
at the considered energies $\chi$ is weakly dependent on $s$, only the latter
possibility remains.

Turning therefore to the case of $LR$ initial longitudinal polarizations, the
analogue of Eq.(\ref{all}) reads:
\begin{equation}{\cal A}^{LR}_{LL}=\frac{s}{M_W^2}\
\left[{\tilde{\cal A}}^{LR}_{LL}+
\left(\Delta k_{\gamma}-\chi\cdot g_Zg_L\Delta k_Z\right)\right],\label{alr}
\end{equation}
where
\begin{equation}{\tilde{\cal A}}^{LR}_{LL}={\tilde{\cal A}}^{LR}_{LL}(\nu)+
{\tilde{\cal A}}^{LR}_{LL}(\gamma,Z),\label{talr}\end{equation}
\begin{equation}{\tilde{\cal A}}^{LR}_{LL}(\gamma,Z)=\frac{3-\beta_W^2}{2}
\left(1-\chi\cdot g_Zg_L\right).\label{talrs}\end{equation}
Referring to the separation of the neutrino exchange amplitude
introduced in Eq.(4), one may notice that only the amplitude $A_{1\nu}$
appears in Eq.(\ref{talr}).
The $LR$ integrated cross section can be obtained similar to Eq.(\ref{sll}),
using the formulae given in the Appendix. Also in this case one finds two
allowed bands for $\Delta k_{\gamma}$ and $\Delta k_Z$. However, taking into
account that $g_Zg_L=-1.17$ (or $-1$ for ${\sin^2\theta_W}=0.25$),
as opposed to $g_Zg_R=+1$, these bands limit the combination
$\vert \Delta k_{\gamma}+1.17\chi\cdot\Delta k_Z\vert$, and are thus almost
orthogonal to the bands previously derived in the $RL$ case for the
combination $\vert \Delta k_{\gamma}-\chi\cdot\Delta k_Z\vert$,
see Eqs.(\ref{k1}) and (\ref{k2}).
Concerning the widths of the
$LR$ and the $RL$ bands, by actual numerical calculations these are found to
be qualitatively of the same size, provided one limits the angular integration
of $\sigma^{LR}$ to a range including the backward hemisphere, for example
the range $-0.9\leq\cos\theta\leq 0.3$ which will be considered later on for
a more quantitative analysis. The lower limit of integration is chosen from
experimental conditions, while the upper one minimizes the ``background''
from the neutrino exchange diagram, which dominates in the forward direction
and does not contain the trilinear coupling constants, thus reducing the
sensitivity in the forward hemisphere [19]. Corresponding to this integration
range, one obtains the bands labelled in Fig.4 as `3'
and `4', which together with `1' and `2' determine four allowed regions for
the coupling constants, respectively labeled as {\it a, b, c, d}.
Since numerically ${\tilde{\cal A}}_{LL}^{LR}(\nu)<0$,
${\tilde{\cal A}}_{LL}^{LR}(\gamma,Z)>0$, with
$\vert{\tilde{\cal A}}_{LL}^{LR}(\nu)\vert >\vert
{\tilde{\cal A}}_{LL}^{LR}(\gamma,Z)\vert$, one can see that
${\tilde{\cal A}}_{LL}^{LR}<0$ (see Eq.(\ref{talr})) shifts the position of
the band `4' upwards with respect to the band `3'. Note that only the
region {\it a} in Fig.4 is compatible with the SM values
$\Delta k_Z=\Delta k_{\gamma}=0$.

While the ``ideal'' situations presented above assumed 100\% $RL$ or $LR$
polarizations,
in practice the cross section is expressed as:
\begin{equation}\sigma=\frac{1}{4}\left[(1+P_1)\cdot (1-P_2)\hskip 2pt
\sigma^{RL}+(1-P_1)\cdot (1+P_2)\hskip 2pt\sigma^{LR}\right],\label{gcr}
\end{equation}
where $P_1$ ($P_2$) are less than unity, and represent the actual degrees of
longitudinal polarization of $e^-$ ($e^+$). For any two different
couples of values for ($P_1,\hskip 2pt P_2$), there will be four
intersections, similar to Fig.4. Thus, by varying
$P_1$ and $P_2$,
the slopes of the bands in Fig.4 change, in such a way that the allowed
region corresponding to {\it a}
remains in the same position (around the origin), and the other intersections
change their positions in the $(\Delta k_{\gamma},\hskip 2pt\Delta k_Z)$ plane,
and therefore should be excluded. Accordingly, one would need to perform
experiments at (at least) two different pairs of suitable values
of $P_1$ and $P_2$, in order to reduce the allowed regions to just region
{\it a}. Actually, one can easily see that the extension of this region
{\it a} would
be minimized by the symmetric choice $P_1=-P_2=P>0$ and $P_2=-P_1=P>0$, if
experimentally feasible.

Combining Fig.4 with Eq.(\ref{atilda}) and the subsequent discussion
on $\sigma^{RL}_{LL}$ and $\sigma^{LR}_{LL}$,
we qualitatively obtain that the expected constraints on the anomalous
coupling constants could be of the order of
$\vert\Delta k_V\vert< 10^{-3}\ (10^{-4})$ for the two considered values of
$\sqrt s$.

As it was mentioned in Section 1, in principle the uncertainty on the
polarizations $P_1$ and $P_2$ can induce a large systematic uncertainty on
the determination
of $\sigma^{RL}$, because $\sigma^{LR}\gg\sigma^{RL}$ for total cross
sections.
The effect of this contamination can be substantially reduced by observing
that the bulk of $\sigma^{LR}$ comes from the forward direction, and that the
ratio between $\sigma^{RL}$ and $\sigma^{LR}$ is not so negligibly small in
the backward direction, say $\cos\theta<0$, as exemplified in Fig.5. This
suggests that one should limit to just this kinematical region, without
substantial loss on the statistics for $\sigma^{RL}$. In this
kinematical region $\sigma^{LR}/\sigma^{RL}=10-20$, so that, with
{\it e.g.} $P_1=-P_2=0.8$, and with $\delta P_1/P_1=\delta P_2/P_2=0.01$,
such an uncertainty on the polarization induces on $\sigma^{RL}$ a
{\it systematic} uncertainty of the order of
${\displaystyle{\left(\delta\sigma/\sigma\right)^{RL}_{sys}\simeq 10^{-2}}}$.
This is to be compared to a statistical uncertainty on the $RL$ cross
section (in the backward region) of about $6\cdot 10^{-2}$ (at
$\sqrt s=0.5\ TeV$).

We turn now to the process of transversely polarized $W^{\pm}$
production for different polarizations of the initial $e^+e^-$ beams, and first
compare its features with the case, considered previously, of
$W^+_L\hskip 2pt W^-_L$ production. In particular,
as noticed above with regard to Fig.3, the corresponding SM cross section
$\sigma^{RL}_{TT}$ at the energies of interest here is too small to provide
sufficient statistics. Indeed,
$\sigma^{RL}_{TT}=0.4\ fb\ (7\cdot 10^{-3}\ fb)$ at
$\sqrt{s}=0.5\ TeV\ (1\ TeV)$.
Consequently, for a one year run we expect only three (less than one) event
samples. On
the other hand, as can be seen from Eq.(\ref{att}), the contributions coming
from
anomalous terms increase the cross section due to the enhancement factor
$s/2M_W^2$ contained in amplitude. Therefore, although the SM cross section
could not be observable at a given level of accuracy, the cross section
with large enough values of the anomalous vertices might be observable
at high energy.
Thus, to the purpose of deriving restrictions on the anomalous couplings from
transversally polarized $W^\pm$ production, we can adopt as a criterion for
observability of $TT$ events, the condition that the relevant
cross section should be observable with an uncertainty
\begin{equation}\delta\sigma^{RL}_{TT}(\lambda_V)<\sigma^{RL}_{TT}(\lambda_V)
-\sigma_{TT}^{RL}(SM)\label{crit}.\end{equation}
The reverse of this inequality will give upper limits on $\lambda_V$. We can
notice that in this case the criterion is different from the one in
Eq.(\ref{delv}), where the SM cross section is assumed to be measured.
Actually, it would conceptually coincide with Eq.(\ref{delv})
if $\sigma(SM)$ was measurable with good
statistics, whereas for $\sigma(SM)\ll\sigma (\lambda_V)$ it requires that
at least four $W^-_T\hskip 2pt W^+_T$ production events are observed in order
to be able to state that $\sigma(\lambda_V)\not=0$ at the 95\% C.L.

The comparison between the amplitudes of process (1) for the production of
longitudinally and transversely polarized $W^\pm$ pairs shows that, due to
their similar dependence on anomalous couplings
(see Eqs.(\ref{all}) and (\ref{att})), the expected bounds on $\lambda_V$
in the $(\lambda_\gamma,\lambda_Z)$ plane should
qualitatively have the same form as the band `1' in Fig.4, previously
derived in the ($W^+_LW^-_L$) case from Eq.(\ref{k1}). Indeed,
from Eq.(\ref{att}) and the above criterion, for the $RL$
case one has (assuming $\sigma^{SM}\ll\sigma(\lambda_V)$) the upper limit
\begin{equation}
\vert\lambda_\gamma-\chi\cdot\lambda_Z\vert<
\frac{4}{\sqrt{(\sigma_0)_{TT}\varepsilon_W L_{int}}}\cdot\frac{M_W^2}{s},
\label{l1}
\end{equation}
which expresses the $95\%$ C.L. bounds. In (\ref{l1}), we denote
$(\sigma_0)_{TT}=4\pi\alpha_{\it e.m.}^2\beta_W^3/3s=8(\sigma_0)_{LL}$.
Contrary to the case of Eqs.(17) and (18), we have here just one inequality,
because we consider only positive sign for the deviation on right side of
Eq.(\ref{crit}).
For the inputs used in the previous cases, the typical values of the right side
of Eq.(\ref{l1}), characterizing the width of the band
in the $(\lambda_\gamma,\lambda_Z)$ plane labeled as `1' in Fig.6,
are $1.8\cdot 10^{-3}$ ($6.5\cdot 10^{-4}$) at
$\sqrt{s}=0.5\ TeV\ (1\ TeV)$.

At this point, analogous to the case studied above of longitudinally
polarized $W^\pm$ bosons, the $LR$ polarization of the
initial $e^-e^+$ beams with final transversely polarized $W$'s could be used
to change the slope of the band obtained from Eq.(\ref{l1}) and turn it into a
band limiting $\vert\lambda_{\gamma}+1.17\chi\cdot\lambda_Z\vert$. A
finite domain allowed to $\lambda_{\gamma}$ and $\lambda_Z$ should then
occur from the combination of the two bands.
However, in contrast to the $W^+_LW^-_L$ case, where the widths of
$RL$ and $LR$ numerically turn out to be comparable, the $\nu$-mediated
amplitude for production $W^+_TW^-_T$ has the additional part $A_{2\nu}$
with $\Delta\lambda=\pm 2$. This amplitude does not
interfere with the $s$-channel amplitudes (see Eq.(4)), and at the same time
significantly increases the SM cross section as well as the cross section
with anomalous couplings. Recalling Eq.(4) and the criterion (\ref{crit}),
one can see that the presence of $A_{2\nu}$ affects the
limits on $\vert\lambda_{\gamma}+1.17\chi\cdot\lambda_Z\vert$ by a factor
$\sqrt{1+(\sigma_2/\sigma_1)}$, where
$\sigma_{2\nu}\propto\vert A_{2\nu}\vert^2$ and
$\sigma_{1\nu}\propto\vert A_{\gamma}+A_{1\nu}+A_Z\vert^2$. Numerically,
$\sigma_{2\nu}$ by far dominates over the $\Delta\lambda=0$ cross section
$\sigma_{1\nu}$, and dramatically increases the factor mentioned above and
the corresponding width of the allowed band for
$\vert\lambda_\gamma+1.17\chi\cdot\lambda_Z\vert$. Only in the specific
kinematical region of outgoing $W^-$ in the backward direction, the ratio of
cross sections (integrated over the backward hemisphere) can reduce to as low
as $\sim 10^2$, but not less. In this case the width of $LR$ band
would be approximately ten times larger than the $RL$ one, as it is shown in
Fig.6. One can see that the width of band `1' allowed by
$\sigma^{RL}_{TT}$ is of the order of $10^{-3}$, whereas band `2'
allowed by $\sigma^{LR}_{TT}$ has a
width of the order of $10^{-2}$. Consequently, we can conclude that the
anomalous couplings ($\lambda_\gamma, \lambda_Z)$
can be strongly correlated, rather than being severely restricted in
a symmetric small region. For $\sqrt s=1\ TeV$, and the corresponding assumed
luminosity, the limits of Fig.6 are improved by a factor
$\sqrt{L_{int}\hskip 2pt s}=\sqrt 8$.

\section{Discussion and concluding remarks}

The integrated and differential cross-sections  as well as the asymmetries
of the process under consideration are commonly considered as the basic
experimental observables to study deviations from the SM induced by new
physical effects. Continuing our previous discussion of the sensitivity
of the $e^+e^-\to W^+W^-$ process to the anomalous $WW\gamma$ and
$WWZ$ couplings, we will use, for the different cases corresponding to
specific initial and final polarizations, the integrated cross section
$\sigma(z_1,z_2)$ and the forward-backward asymmetry $A_{FB}$, respectively
defined as ($z\equiv\cos\theta$):
\begin{equation}\sigma(z_1,z_2)=\int_{z_1}^{z_2}\frac{d\sigma}{dz}dz;
\qquad\quad A_{FB}=\frac{\sigma(0,z_2)-\sigma(z_1,0)}{\sigma(z_1,z_2)}
\hskip 2pt .
\label{afb}\end{equation}
To work out an example closer to the realistic situation, which somehow
could account for both statistical and
systematic experimental uncertainties, we assume a systematic error in the
cross-section measurement at the level of $\sim$2\% [28], resulting from
an uncertainty in the luminosity measurement ($\delta L_{int}\simeq$1\%),
an error in the acceptance ($\delta_{accept}\simeq$1\%),
an error for background subtraction ($\delta_{backgr}\simeq$0.5\%),
a systematic error on the knowledge of the branching ratio of
$W\rightarrow {\bar f}f$ ($\delta_{Br}\simeq$0.5\%). Finally, we assume the
degrees of longitudinal polarizations $\vert P_1\vert,\hskip 2pt
\vert P_2\vert = 0.8$, with an
uncertainty $\delta{P_1}/P_1=\delta{P_2}/P_2=10^{-2}$. We notice that for the
integrated luminosities assumed in the previous Section, $L_{int}=50\ fb^{-1}$
and $100\ fb^{-1}$ at $\sqrt s=0.5\ TeV$ and $1\ TeV$ respectively, the
systematic uncertainty would dominate over the statistical one in the case
of initial $LR$ and unpolarized $e^-e^+$, whereas the converse would be true
for
initial $RL$ due to the smallness of this cross section. The numerical results
for the bounds on the anomalous couplings presented in this Section are derived
using a combined $\chi^2$ analysis of the integrated cross sections and of
the forward-backward asymmetry. All allowed domains will be given
to 95\% C.L., corresponding to $\Delta\chi^2=5.99$ for two
simultaneously fit free parameters.

In general, we have four free couplings to fit to the experimental
data. For unpolarized initial and final states, it is therefore not possible
to disentangle the dependence of the cross section on the various constants.
The simplest procedure is to fix a couple of parameters at the SM values, and
derive allowed regions for the remaining ones. In Figs.7a,b we fix
$\lambda_Z,\hskip 3pt k_Z$, and consider cross sections at $\sqrt s=0.5\ TeV$,
integrated in the ``optimal'' range $-0.9\leq z\leq 0.3$ already
introduced in the previous Section, where anomalous effects are most
pronounced.
The regions allowed to $\lambda_{\gamma},\hskip 3pt k_{\gamma}$
by such a procedure are the ones
enclosed by the dashed ellipses in Fig.7a for unpolarized beams, while
initial longitudinal $e^-e^+$ ``$LR$''
polarization with $P_1=-P_2=-0.8$ provides the allowed region enclosed by
the bigger full ellipses. The intersection of the former region with the latter
one already restricts the allowed domain. Furthermore, initial
$e^-e^+$ ``$RL$''
polarization with $P_1=-P_2=0.8$ gives the area enclosed by the smaller
solid ellipses in Fig.7a, thus further (and quite significantly)
restricting the range of allowed values for the coupling constants
$\lambda_{\gamma},\hskip 3pt k_{\gamma}$ to the shaded region. This region is
magnified in Fig.7b. In Fig.8 we show the results of the same
analysis, for $\sqrt s=1\ TeV$.

In Figs.9a,b  we fix, instead, $\lambda_{\gamma}$ and $k_{\gamma}$ at the SM
values, referring again to $\sqrt s=0.5\ TeV$. The domain allowed to
$\lambda_Z,\hskip 3pt k_Z$ by the unpolarized
cross section is the interior of the dashed ellipses, while the
initial ``$LR$'' and ``$RL$'' polarizations
give the regions enclosed by the bigger and by the smaller solid ellipses,
respectively. Analogous to Fig.7a, the shaded restricted domain represents
the values of $\lambda_Z,\hskip 3pt k_Z$ allowed by the combination of
polarized cross sections. This domain is magnified in Fig.9b. Furthermore,
in Fig.10 we repeat the same analysis, for $\sqrt s=1\ TeV$.

We now consider the case of both initial $e^+e^-$ and final $W^+W^-$
polarizations, where in principle one may attempt to disentangle the
anomalous coupling pairs $(k_{\gamma},\hskip 2pt k_Z)$ and
$(\lambda_{\gamma},\hskip 2pt \lambda_Z)$.
Under the same input conditions leading to Figs.7--10, one finds the allowed
regions displayed in Figs.11-12 for both $W^{\pm}$ longitudinally polarized,
for the CM energies $\sqrt s=0.5$ and $1\ TeV$.
Specifically, the bands labeled as `3' represent the regions allowed by
unpolarized initial $e^-e^+$ beams, the dashed bands `1' are allowed by initial
``$RL$'' polarization, and the dashed bands `2' are the ones allowed by
``$LR$'' polarization. As one can see, the combined measurements lead to a
restricted area for the
values of $(k_{\gamma},\hskip 2pt k_Z)$, and the most restrictive case
corresponds to the combination of ``$LR$'' with ``$RL$''. The typical
values of the bounds can be read from Fig.11-12, and are of the same order of
magnitude as those indicated in the previous Section. This should be a
convincing example of the essential role played by measurements of cross
sections with
{\it initial} beams polarization in improving the bounds on anomalous
three-boson vertices obtainable from process (1). In particular,
for $(\lambda_{\gamma},\hskip 2pt \lambda_Z)$, one
needs measurements of cross sections with both $W^{\pm}$ transversely
polarized, and the opportunities are qualitatively the same as the ones
depicted in Fig.6, and discussed at the end of
the previous Section.

The situation of the bounds would not be dramatically different from the one
presented above, in the case initial longitudinal polarization was available
only for the electron beam, the positron beam being unpolarized. For a
right-handed electron beam ($e^-_R$), the cross section would be
$\sigma^R=\sigma^{RL}/2$,
and the corresponding bounds would be higher by a factor $\sqrt 2$, due to
the fact that in this case the dominant uncertainty is the statistical one.
For a left-handed electron beam ($e^-_L$), we expect the bounds to remain
almost the same as presented above, because in this case the systematic
uncertainty is dominating.

Finally, we can remark that the bounds reach the order of
magnitude at which anomalous vector boson couplings appear as the result
of one-loop corrections [29]. In this case one should include such corrections
in the cross sections appearing in Eq.(16), in order to evidence non-standard
contributions to the trilinear vector boson couplings.

\section*{Acknowledgements}

\noindent One of the authors (A.A.P.) acknowledges the support of INFN-Sezione
di Trieste. He also
thanks Prof. Abdus Salam, the International Atomic Energy Agency and
UNESCO for support and hospitality at the International Centre for
Theoretical Physics, Trieste.

\newpage
\section*{Appendix}
\appendix
The cross section of process (1) for arbitrary degrees of longitudinal
polarization of electrons ($P_1$)
and positrons ($P_2$) are  generally expressed by Eq.(\ref{gcr}).
The corresponding polarized differential cross sections
of the process $e^+_be^-_a\to W^+_\beta W^-_{\alpha}$
contained in Eq.(\ref{gcr})
can be written
as follows:
\begin{equation}
\frac{d\sigma^{ab}_{\alpha\beta}}{d\cos\theta}=
C\cdot \sum_{i=0}^{i=11} F_i^{ab}{\cal O}_{i\ \alpha\beta},\label{A1}
\end{equation}
where $C=\pi\alpha^2_{e.m.}\beta_W/2s$, the helicities of the initial
$e^-e^+$ and final $W^-W^+$ states are labeled as $ab=(RL,\ LR)$
and $\alpha\beta=(LL,\ ,TT,\ TL)$, respectively.
In Eq.(\ref{A1}) we follow the notation used in Ref.[30]. In particular, the
${\cal O}_i$ are functions of the kinematical variables which characterize
to the various possibilities for the final $W^+W^-$ polarizations
($TT,\ LL,\ TL+LT$ or the sum of all $W^+W^-$ polarization states for
unpolarized W's). The $F_i$ are combinations of coupling constants including
the anomalous trilinear self-couplings of $W$ bosons.

For the $RL$ case we have:
\begin{eqnarray}
\label{A2}
F_1^{RL} & = & 2(1-g_Zg_R\cdot\chi)^2 \nonumber \\
F_3^{RL} & = & \Delta k_{\gamma}-g_Zg_R(\Delta k_{\gamma}+
\Delta k_Z)\cdot\chi+(g_Zg_R\cdot\chi)^2\Delta k_Z \nonumber \\
F_4^{RL} & = & \lambda_{\gamma}-g_Zg_R(\lambda_\gamma+
\lambda_Z)\cdot\chi+(g_Zg_R\cdot\chi)^2\lambda_Z \nonumber \\
F_9^{RL} & = & \frac{1}{2}(\Delta k_\gamma-g_Zg_R\Delta k_Z\cdot\chi)^2
\nonumber \\
F_{10}^{RL} & = & \frac{1}{2}(\lambda_\gamma-g_Zg_R\lambda_Z\cdot\chi)^2
\nonumber \\
F_{11}^{RL} & = & \frac{1}{2}\left[\Delta k_\gamma\lambda_{\gamma}-
g_Zg_R(\Delta k_\gamma\lambda_Z+\Delta k_Z\lambda_\gamma)\cdot\chi+
(g_Zg_R\cdot\chi)^2\Delta k_Z\lambda_Z\right]
\end{eqnarray}

The remaining $F^{RL}$ are zero.

For the $LR$ case we have:
\begin{eqnarray}
\label{A3}
F_0^{LR} & = & \frac{1}{16s^4_W} \nonumber \\
F_1^{LR} & = & 2(1-g_Zg_L\cdot\chi)^2 \nonumber \\
F_2^{LR} & = & -\frac{1}{2s^2_W}(1-g_Zg_L\cdot\chi) \nonumber \\
F_3^{LR} & = & \Delta k_{\gamma}-g_Zg_L(\Delta k_{\gamma}+
\Delta k_Z)\cdot\chi+(g_Zg_L\cdot\chi)^2\Delta k_Z \nonumber \\
F_4^{LR} & = & \lambda_{\gamma}-g_Zg_L(\lambda_\gamma+
\lambda_Z)\cdot\chi+(g_Zg_L\cdot\chi)^2\lambda_Z \nonumber \\
F_6^{LR} & = & -\frac{1}{4s^2_W}(\Delta k_\gamma-g_Zg_L\Delta k_Z\cdot\chi)
\nonumber \\
F_7^{LR} & = & -\frac{1}{4s^2_W}(\lambda_\gamma-g_Zg_L\lambda_Z\cdot\chi)
\nonumber \\
F_9^{LR} & = & \frac{1}{2}(\Delta k_\gamma-g_Zg_L\Delta k_Z\cdot\chi)^2
\nonumber \\
F_{10}^{LR} & = & \frac{1}{2}(\lambda_\gamma-g_Zg_L\lambda_Z\cdot\chi)^2
\nonumber \\
F_{11}^{LR} & = & \frac{1}{2}\left[\Delta k_\gamma\lambda_{\gamma}-
g_Zg_L(\Delta k_\gamma\lambda_Z+\Delta k_Z\lambda_\gamma)\cdot\chi+
(g_Zg_L\cdot\chi)^2\Delta k_Z\lambda_Z\right]
\end{eqnarray}
The remaining $F^{LR}$ are zero. In (\ref{A2}) and (\ref{A3})
$s^2_W\equiv{\sin^2\theta}_W$; $g_{L},g_{R}=v\pm a$, where $v$ and $a$ are
the vector and axial-vector $Ze^+e^-$ couplings ($c_W\equiv\cos\theta_W$):
\begin{equation}
v=\frac{T_3^e-2Q_e s^2_W}{2s_Wc_W}, \;\;\;
a=\frac{T_3^e}{2s_Wc_W}.
\end{equation}
Eqs.(\ref{A2})--(\ref{A3}) are obtained in the approximation where
the imaginary part of $Z$ boson propagator is neglected. Accounting for
this  effect requires the replacements $\chi\to Re\chi$ and
$\chi^2\to\vert\chi\vert^2$ in right-hand sides of Eqs.(\ref{A2})--(\ref{A3}).

In Eq.(\ref{A1}), for the longitudinal (LL) cross sections
${\displaystyle{\frac{d\sigma(e^+e^-\to W^+_LW^-_L)}{d\cos\theta}}}$ we have
(with $\vert\vec p\vert=\sqrt{s}\beta_W/2$):
\begin{eqnarray}
{\cal O}_{0,LL} & = & \frac{s(1-{\cos^2\theta})}{4t^2M^4_W}\left[s^3(1+
{\cos^2\theta})-4M_W^4(3s+4M_W^2)-
4(s+2M_W^2)\vert\vec p\vert s\sqrt{s}\cos\theta\right] \nonumber \\
{\cal O}_{1,LL} & = & \frac{s^3-12sM_W^4-16M_W^6}{8sM_W^4}(1-{\cos^2\theta})
\nonumber \\
{\cal O}_{2,LL} & = & \frac{1-{\cos^2\theta}}{t}\left[\frac{\vert\vec p\vert s
\sqrt {s}(s+2M_W^2)}{2M_W^4}\cos\theta-\frac{s^3-12sM_W^4-16M_W^6}
{4M_W^4}\right]\nonumber \\
{\cal O}_{3,LL} & = & \frac{s^2-2M_W^2s-8M_W^4}{2M^4_W}(1-{\cos^2\theta})
\nonumber \\
{\cal O}_{4,LL} & = & {\cal O}_{5,LL}={\cal O}_{7,LL}={\cal O}_{8,LL}=
{\cal O}_{10,LL}={\cal O}_{11,LL} \nonumber \\
{\cal O}_{6,LL} & = & \frac{s(1-{\cos^2\theta})}{2tM_W^4}\left[8M_W^4+
2sM_W^2-s^2+2s\vert\vec p\vert\sqrt{s}\cos\theta\right] \nonumber \\
{\cal O}_{9,LL} & = & 2\frac{s\vert\vec p\vert^2}{M_W^4}(1-{\cos^2\theta})
\end{eqnarray}

For the transverse (TT) cross sections
${\displaystyle{\frac{d\sigma(e^+e^-\to W^+_TW^-_T)}{d\cos\theta}}}$ we have:
\begin{eqnarray}
{\cal O}_{0,TT} & = & \frac{4s}{t^2}\left[s(1+{\cos^2\theta})-2M_W^2-
2\vert\vec p\vert\sqrt{s}\cos\theta\right](1-{\cos^2\theta}) \nonumber \\
{\cal O}_{1,TT} & = & \frac{4{\vert\vec p\vert}^2}{s}(1-{\cos^2\theta})
\nonumber \\
{\cal O}_{2,TT} & = & \frac{1-{\cos^2\theta}}{t}\left[4\vert\vec p\vert
\sqrt{s}\cos\theta-8{\vert\vec p\vert}^2\right] \nonumber \\
{\cal O}_{3,TT} & = & {\cal O}_{5,TT}={\cal O}_{6,TT}={\cal O}_{8,TT}=
{\cal O}_{9,TT}={\cal O}_{11,TT} \nonumber \\
{\cal O}_{4,TT} & = & \frac{8{\vert\vec p\vert}^2}{M_W^2}(1-{\cos^2\theta})
\nonumber \\
{\cal O}_{7,TT} & = & \frac{s}{M_W^2}{\cal {O}}_{2,TT} \nonumber \\
{\cal O}_{10,TT} & = & \frac{4s{\vert\vec p\vert}^2}{M_W^4}
(1-{\cos^2\theta})
\end{eqnarray}

Finally, for the production of the one longitudinal plus one transverse
vector boson $(TL+LT)$ we have:

\begin{eqnarray}
{\cal O}_{0,TL} & = & \frac{2s}{t^2M^2_W}\left[s^2(1+\cos^4\theta)-
4\vert\vec p\vert\sqrt{s}\ cos\theta(4{\vert\vec p\vert}^2+
s\cos^2\theta)+\right. \nonumber \\
  & & \left. 4M_W^4(1+\cos^2\theta)+2s(s-6M_W^2)\cos^2\theta
-4sM_W^2 \right] \nonumber \\
{\cal O}_{1,TL} & = & \frac{4{\vert\vec p\vert}^2}{M_W^2}(1+{\cos^2\theta})
\nonumber \\
{\cal O}_{2,TL} & = & {\cal O}_{6,TL}={\cal O}_{7,TL}=
\frac{4\vert\vec p\vert\sqrt{s}}{tM_W^2}\left[(4{\vert\vec p\vert}^2+
s\cos^2\theta)\cos\theta-
2{\vert\vec p\vert}\sqrt{s}(1+\cos^2\theta)\right]\nonumber \\
{\cal O}_{3,TL} & = & {\cal O}_{4,TL}={\cal O}_{11,TL}=2{\cal O}_{9,TL}=
2{\cal O}_{10,TL}=\frac{8{\vert\vec p\vert}^2}{M_W^2}(1+
\cos^2\theta)  \nonumber \\
{\cal O}_{5,TL} & = & \frac{32{\vert\vec p\vert}^3\sqrt{s}}{M_W^4}
\cos\theta \nonumber \\
{\cal O}_{8,TL} & = & \frac{16s{\vert\vec p\vert}^2}{tM_W^4}
\left[M_W^2+2\vert\vec p\vert\sqrt{s}\cos\theta-
(s-M_W^2)\cos^2\theta\right]
\end{eqnarray}

\newpage
\section*{Figure captions}
\begin{description}
\item[1.] The $WWV$ vertex $(V=\gamma ,\: Z)$.

\item[2.] The Feynman diagrams for the $e^+e^-\to W^+W^-$.

\item[3.] The energy behaviour of the total SM cross sections for
$e^-_Re^+_L\to W^-W^+$ (thin solid curve), $e^-_Re^+_L\to W^-_LW^+_L$ (dotted
curve) and $e^-_Re^+_L\to W^-_TW^+_T$ (thick solid curve).

\item[4.] Qualitative features of allowed regions for
($\Delta k_{\gamma},\; \Delta k_Z$) from $e^-_Re^+_L\to W^-_LW^+_L$
(bands `1' and `2') and $e^-_Le^+_R\to W^-_LW^+_L$ (bands `3' and `4').
$\delta$ is defined as: ${\displaystyle{\frac{1}{2}\left(\frac{\delta\sigma}
{\sigma}\vert\tilde{\cal A}\vert\right)^{RL}_{LL}}}$.

\item[5.] Differential SM cross section for
$e^+e^-\to W^+W^-$ at
$\sqrt s=500\ GeV$ for $e^-e^+$ unpolarized (thin solid curve),
$LR$ (dotted curve) and $RL$ (thick solid curve).

\item[6.] Allowed regions (95\% C.L.) for
($\lambda_{\gamma},\; \lambda_Z$) from $e^-_Re^+_L\to W^-_TW^+_T$
(band `1') and $e^-_Le^+_R\to W^-_LW^+_L$ (band `2') at $\sqrt s=500\ GeV$.

\item[7a.] Allowed domains (95\% C.L.) for ($k_{\gamma},\: \lambda_{\gamma}$)
for fixed $\lambda_Z=\Delta k_Z=0$; $\sqrt s=0.5\ TeV$; $L_{int}=50\ fb^{-1}$;
$\varepsilon_W=0.15$; $-0.9\leq\cos\theta\leq 0.3$.
Smaller solid ellipses: $P_1=-P_2=0.8$; dashed ellipses: $P_1=P_2=0$;
bigger full ellipses: $P_1=-P_2=-0.8$. Shaded allowed area: combination of
polarized cross sections.

\item[7b.] Same as Fig.7a, magnified allowed domain.

\item[8.] Same as Fig.7b, for $\sqrt s=1\ TeV$; $L_{int}=100\ fb^{-1}$;
$\varepsilon_W=0.15$; $-0.9\leq\cos\theta\leq 0.3$.

\item[9a.]  Similar to Fig.7a, allowed bounds (95\% C.L.) for
($k_Z$, $\lambda_Z$) for fixed $\lambda_\gamma=\Delta k_\gamma=0$.

\item[9b] Same as Fig.9a, magnified allowed domain.

\item[10.] Same as Fig.9b, for $\sqrt s=1\ TeV$; $L_{int}=100\ fb^{-1}$;
$\varepsilon_W=0.15$; $-0.9\leq\cos\theta\leq 0.3$.

\item[11.] Allowed domains (95\% C.L.) for
($k_{\gamma},\; k_Z$) from $e^-e^+\to W^-_LW^+_L$; $\sqrt s=0.5\ TeV$;
$L_{int}=50\ fb^{-1}$; $\varepsilon_W=0.15$;
$-0.9\leq\cos\theta\leq 0.3$. Band `1': $P_1=-P_2=0.8$;
band `2': $P_1=-P_2=-0.8$; band `3': $P_1=P_2=0$.
Shaded allowed area: combination of polarized cross sections.

\item[12.] Same as in Fig.11, for $\sqrt s=1\ TeV$;
$L_{int}=100\ fb^{-1}$.

\end{description}

\end{document}